%% file: skeleton.tex
\documentclass[a4paper,11pt]{article}
\usepackage{pos}
\usepackage{amsmath}
\usepackage{bbold}
\usepackage{cancel}
\usepackage{xcolor}
\usepackage{comment}
\usepackage{caption}
\usepackage{graphicx}
\usepackage{subcaption}
\usepackage{hyperref}
\usepackage[normalem]{ulem}

\title{Charm quark mass using a massive nonperturbative renormalisation scheme}
\ShortTitle{Charm quark mass using a massive NPR scheme}

\author[a]{Luigi~Del~Debbio}
\author[b]{Felix~Erben}
\author[c]{Jonathan~Flynn}
\author*[c]{Rajnandini~Mukherjee}
\author[b]{J.~Tobias~Tsang}

\affiliation[a]{Higgs Centre for Theoretical Physics, University of Edinburgh, Edinburgh, UK}
\affiliation[b]{CERN, Theoretical Physics Department, Geneva, Switzerland}
\affiliation[c]{School of Physics and Astronomy, University of Southampton, Southampton, UK}

\emailAdd{r.mukherjee@soton.ac.uk}

\abstract{We present a first numerical implementation of a massive nonperturbative renormalisation scheme, RI/mSMOM, in the study of heavy quarks using the domain-wall fermion action. In particular, we calculate renormalisation constants for fermion bilinears at non-vanishing heavy-quark masses and compare the approach to the continuum of the renormalised charm-quark mass with that from a mass-independent scheme.}

\renewcommand{\hookAfterAbstract}{%
\par\bigskip
\textsc{preprint}: CERN-TH-2023-250}

\FullConference{%
  The 39th International Symposium on Lattice Field Theory\\
  31 July - 4 August 2023\\
  Fermilab, Batavia, Illinois
}

\begin{document}
\maketitle

\input{1_introduction}
\input{2_NPR}
\input{3_comp_details}
\input{4_method}

\input{5_results}

\input{6_conclusions}

\input{7_acknowledgements}

\bibliographystyle{JHEPmod}
\bibliography{msmom.bib}

\end{document}

%% file: 1_introduction.tex

\section{Introduction}
Lattice simulations allow for first principles computations of Standard Model parameters such as quark masses in the nonperturbative regime of QCD. However, the bare quantities calculated on the lattice need to be renormalised using some intermediate renormalisation scheme for them to have a well-defined continuum limit. There are many such schemes for the nonperturbative renormalisation (NPR) of bare lattice quantities, such as RI/MOM \cite{Martinelli_1995}, RI/SMOM \cite{Sturm:2009kb} and the Schr\"{o}dinger functional method \cite{Luscher:1992an,Sint:1993un}. However, these schemes are implemented in the chiral limit of QCD, and therefore introduce $am$-sized lattice artefacts when used for renormalising heavy-quark observables. A massive momentum-subtraction scheme, called RI/mSMOM, prescribing NPR away from the chiral limit, has been introduced in \cite{Boyle:2016wis} to possibly ameliorate these cutoff effects. Designed to have similar properties to the RI/SMOM scheme (which is already $O(am)$-improved), this massive scheme aims to change $O(a^2m^2)$-sized lattice artefacts in heavy-quark observables, potentially leading to smoother continuum extrapolations. 

In this study, we present a pilot numerical implementation of the RI/mSMOM (or \textit{massive}) scheme on three ensembles with three different lattice spacings, including a comparison to the RI/SMOM (or \textit{massless}) scheme. We investigate its effectiveness in mitigating cutoff effects by studying in particular the renormalised charm-quark mass.

%% file: 2_NPR.tex

\section{Nonperturbative renormalisation}

\begin{figure}
    \centering
    \includegraphics{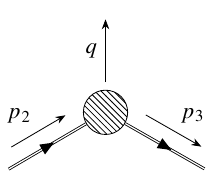}
    \caption{Choice of kinematics. The dashed bubble represents the operator insertion and higher order corrections, $p_2$ and $p_3$ are the momenta of the external off-shell quark lines.}
    \label{fig:kinematics}
\end{figure}

Let us consider the Green's function of a quark bilinear operator $\mathcal{O}_\Gamma = \overline{\psi}_f\Gamma \psi_{f'}$ between two external off-shell quark lines in a fixed (Landau) gauge, given by (all quantities in Euclidean space)
\begin{align}
  G_\Gamma(p_3,p_2) = \langle\psi_{f}(p_3) \mathcal{O}_\Gamma(q)\overline{\psi}_{f'}(p_2)\rangle,
\end{align}
where $\psi_f$ and $\psi_{f'}$ are quark fields of different flavours, and $q=p_2-p_3$ (see figure  \ref{fig:kinematics} for conventions used). We are interested in scalar ($\Gamma=\mathbb{1} $), pseudoscalar ($\Gamma=i\gamma_5$), vector ($\Gamma=\gamma^\mu$) and axial vector ($\Gamma=\gamma^\mu\gamma_5$) bilinears. The quark propagator is defined as
\begin{align}
    S_f(p) =  \langle\psi_f (p)\overline{\psi}_f(p)\rangle = \frac{1}{i\cancel{p}+m_f},
\end{align}
and the amputated Green's function is obtained by amputating each leg with the inverse quark propagator of the corresponding flavour
\begin{align}
  \Lambda_{\Gamma}(p_2,p_3) = S^{-1}_f(p_3)G_\Gamma(p_3, p_2)S^{-1}_{f'}(p_2).
\end{align}
In our study, we consider the flavour diagonal case $\psi_f=\psi_{f'} = c$, so we drop the flavour subscripts from here on. The propagators are also related to the amputated Green's functions of the vector and axial vector operators via the Ward-Takahashi identities
\begin{align}
    q_\mu\Lambda_V^\mu(p_2,p_3) &= iS^{-1}(p_3) - iS^{-1}(p_2), \label{eq:WI_V}\\
    q_\mu\Lambda_A^\mu(p_2,p_3) &= -2m\Lambda_P(p_2,p_3) + i\gamma_5S^{-1}(p_2) +  S^{-1}(p_3)i\gamma_5. \label{eq:WI_A}
\end{align}

Renormalised and bare quantities are related via renormalisation constants $Z$ as
\begin{align}
    \psi_R &= Z_q^{1/2}\psi, \quad m_R = Z_mm, \quad \mathcal{O}_{\Gamma,R} = Z_\Gamma\mathcal{O}_\Gamma, \\
    \implies S_R(p) &= Z_qS(p), \quad \Lambda_{\Gamma,R}(p_2,p_3) = \frac{Z_\Gamma}{Z_q}\Lambda_\Gamma(p_2,p_3).
\end{align}
Renormalised quantities are denoted with a subscript $R$, while bare quantities are without a subscript. The renormalisation conditions in the RI/mSMOM scheme are
\begin{align}
  Z_q:\quad &\left.\lim _{m_R \rightarrow \overline{m}} \frac{1}{12 p^2} \operatorname{Tr}\left[-i S_R(p)^{-1} \cancel{p}\right]\right|_{p^2=\mu^2}=1, \label{eq:Z_q}\\
  Z_m:\quad & \lim _{m_R \rightarrow \overline{m}} \frac{1}{12 m_R}\left\{\left.\operatorname{Tr}\left[S_R(p)^{-1}\right]\right|_{p^2=\mu^2}+\left.\frac{1}{2} \operatorname{Tr}\left[\left(iq \cdot \Lambda_{\mathrm{A}, R}\right) \gamma_5\right]\right|_{\mathrm{sym}}\right\}=1, \label{eq:Z_m}\\
  Z_V:\quad & \lim _{m_R \rightarrow \overline{m}} \frac{1}{12 q^2} \operatorname{Tr}\left[\left.\left(q \cdot \Lambda_{\mathrm{V}, R}\right) \cancel{q}\right]\right|_{\mathrm{sym}}=1, \label{eq:Z_V}\\
  Z_A:\quad & \left.\lim _{m_R \rightarrow \overline{m}} \frac{1}{12 q^2} \operatorname{Tr}\left[\left(q \cdot \Lambda_{\mathrm{A}, R}\, +\, 2m_R\Lambda_{P,R}\right) \gamma_5 \cancel{q}\right]\right|_{\mathrm{sym}}=1, \label{eq:Z_A}\\
  Z_P:\quad & \left.\lim _{m_R \rightarrow \overline{m}} \frac{1}{12i} \operatorname{Tr}\left[\Lambda_{\mathrm{P}, R} \gamma_5\right]\right|_{\mathrm{sym}}=1, \label{eq:Z_P}\\
  Z_S:\quad & \left.\lim _{m_R \rightarrow \overline{m}} \left\{\frac{1}{12} \operatorname{Tr}\left[\Lambda_{\mathrm{S}, R}\right] +\, \frac{1}{6q^2}\operatorname{Tr}\left[2m_R\Lambda_{P,R}\gamma_5\cancel{q}\right]\right\}\right|_{\mathrm{sym}}=1. \label{eq:Z_S}
\end{align}
The subscript `sym' denotes the symmetric momentum configuration $q^2=p_2^2=p_3^3=\mu^2$, where $\mu$ is the renormalisation scale. These kinematics are the same as those in the massless scheme, RI/SMOM. The massive scheme requires the introduction of another scale, $\overline m$, a renormalised mass at which the renormalisation conditions are imposed. The massless scheme is recovered in the limit $\overline{m}\to 0$.

These conditions are defined such that the vector and axial vector Ward-Takahashi identities (\ref{eq:WI_V}) and (\ref{eq:WI_A}) are satisfied by the renormalised quantities, and the renormalisation constants preserve the relations
\begin{align}
    Z_V = Z_A = 1, \quad Z_P = 1/Z_m, \quad Z_S=Z_P,
\end{align}
as in the continuum $\overline{\text{MS}}$ scheme.

%% file: 3_comp_details.tex

\section{Computational details}
Given the framework outlined above, the bare quantities needed from lattice simulations are the Green's functions for the bilinear operators $G_\Gamma(p_2,p_3)$ for $\Gamma=S,P,V,A$, and the corresponding external leg propagators $S(p_2)$ and $S(p_3)$. Using these, we can compute the amputated Green's functions and impose the renormalisation conditions to calculate the $Z$-factors in the RI/mSMOM scheme.

We use three RBC/UKQCD ensembles with unphysical pion masses using $N_f=2+1$ flavours of domain wall fermions (DWF). These ensembles, called C1, M1 and F1S~\cite{RBC-UKQCD:2008mhs,RBC:2010qam,Boyle:2017jwu,RBC:2014ntl,Boyle:2018knm}, have inverse lattice spacings $a^{-1} \approx 1.78, 2.38, $ and $2.79\,\mathrm{GeV}$ respectively. We compute bilinear Green's functions and external leg propagators for various bare quark masses $am_0$ including the light-quark mass and heavy-quark masses around the charm-quark mass. This is done for various external momenta $ap_\mu$ corresponding to physical momenta in the range $2\,\mathrm{GeV} \lesssim \sqrt{p^2} \lesssim 4\,\mathrm{GeV}$. 

The key point of the RI/mSMOM scheme is the imposition of the renormalisation conditions at some finite value of a renormalised mass $\overline{m}$. We generate  connected pseudoscalar-pseudoscalar two-point correlation functions for the various choices of simulated bare quark masses. We denote their respective ground-state masses $\eta_h(am)$. Interpolations in the values for $\eta_h$ to a common value on all ensembles can then used to fix $\overline{m}$.

Furthermore, in the DWF setup the bare mass requires additive renormalisation that depends on the domain wall parameters, in addition to multiplicative renormalisation. The renormalised mass is thus $m_R = Z_mm = Z_m(m_0 + m_\text{res})$, where $m_0$ is the bare mass which is an input to the simulation, and the residual mass $m_\text{res}$ is measured for each choice of bare mass. In the discussion that follows, $m$ will always denote the multiplicatively renormalised bare mass.

%% file: 4_method.tex

\section{Methodology}
In this section, we describe the procedure used to implement the RI/mSMOM scheme to the study of the charm-quark mass. As we repeat this calculation on multiple ensembles, we need a lattice-independent reference quantity that allows us to compare the results across different lattice spacings. The $\eta_h$ meson mass in physical units enters as this reference quantity and we use it to set the various scales involved in this procedure.

We start by using the bilinear Green's functions and external leg propagators to calculate the renormalisation constants $Z(a\mu, am)$ in the massive scheme over a range of renormalisation scales $a\mu$ and bare masses $am$, on each ensemble.

Next, we compute the bare charm-quark mass $m_c$ on each ensemble. This is done by interpolating to the bare mass that corresponds to the physical $\eta_c$ mass as listed in the PDG \cite{PDG} as
\vspace{-1mm}
\begin{align}
    m_c = a^{-1}\cdot(am_c) = a^{-1}\cdot\big(am(M_{\eta_h}=M_{\eta_c})\big).
    \label{eq:m_c_bare}
\end{align}

Having computed the bare charm-quark mass, we renormalise it using $Z_m$ as calculated using eqn (\ref{eq:Z_m}). For this we choose the renormalisation scale $a\mu = a\cdot(2$ GeV$)$ (keeping in mind the Rome-Southampton window $\Lambda_\text{QCD}^2\ll \mu^2 \ll (\pi/a)^2$). Additionally, we introduce a physical scale by choosing a value of the $\eta_h$ mass, denoted by $M^\star$. On each ensemble, we can then interpolate to the corresponding bare mass $am^\star$ given by
\vspace{-1mm}
\begin{align}
    am^\star = am(M_{\eta_h}=M^\star).
\end{align}
This choice of a physical $\eta_h$ mass connects the choice of bare masses across the different lattice spacings and allows us to subsequently take a continuum limit to obtain the renormalised charm-quark mass
\vspace{-1mm}
\begin{align}
    m_{c,R}(\mu, M^\star) = \lim_{a\to 0} Z_m(a\mu, am^\star)\cdot a^{-1}\cdot (am_c).
    \label{eq:m_c_ren}
\end{align}

The renormalised mass $\overline{m}$ is calculated by extrapolating the renormalised bare mass $m^\star$
\begin{align}
    \overline{m}(\mu, M^\star) = \lim_{a\to 0} Z_m(a\mu, am^\star)\cdot a^{-1}\cdot (am^\star).
    \label{eq:mbar}
\end{align}

The choice of the physical scale $M^\star$ therefore gives us a handle on the scheme-defining renormalised mass $\overline{m}$. We perform this continuum extrapolation for different choices of $M^\star$ --- corresponding to different $\overline{m}$ --- to study the absorption of cutoff effects in the final results, and compare it to the massless scheme. In order to make this comparision, the final step of the procedure is to match the results for each $\overline{m}$ to a scheme $S$ in the continuum (for example, $\overline{\text{MS}}$) such that the dependence on $\overline{m}$ is removed
\begin{align}
    m_{c,R}^S(\mu) = R_m^{S\leftarrow \text{mSMOM}}\left(\mu,\overline{m}\right)m_{c,R}(\mu,\overline{m}).
\end{align}
It is expected that the renormalised charm-quark mass from the different choices of $\overline{m}$ should coincide with each other and with results from RI/SMOM in this definition.

%% file: 5_results.tex

\section{Results}
\begin{figure}
    \centering
    \includegraphics[width=0.82\textwidth]{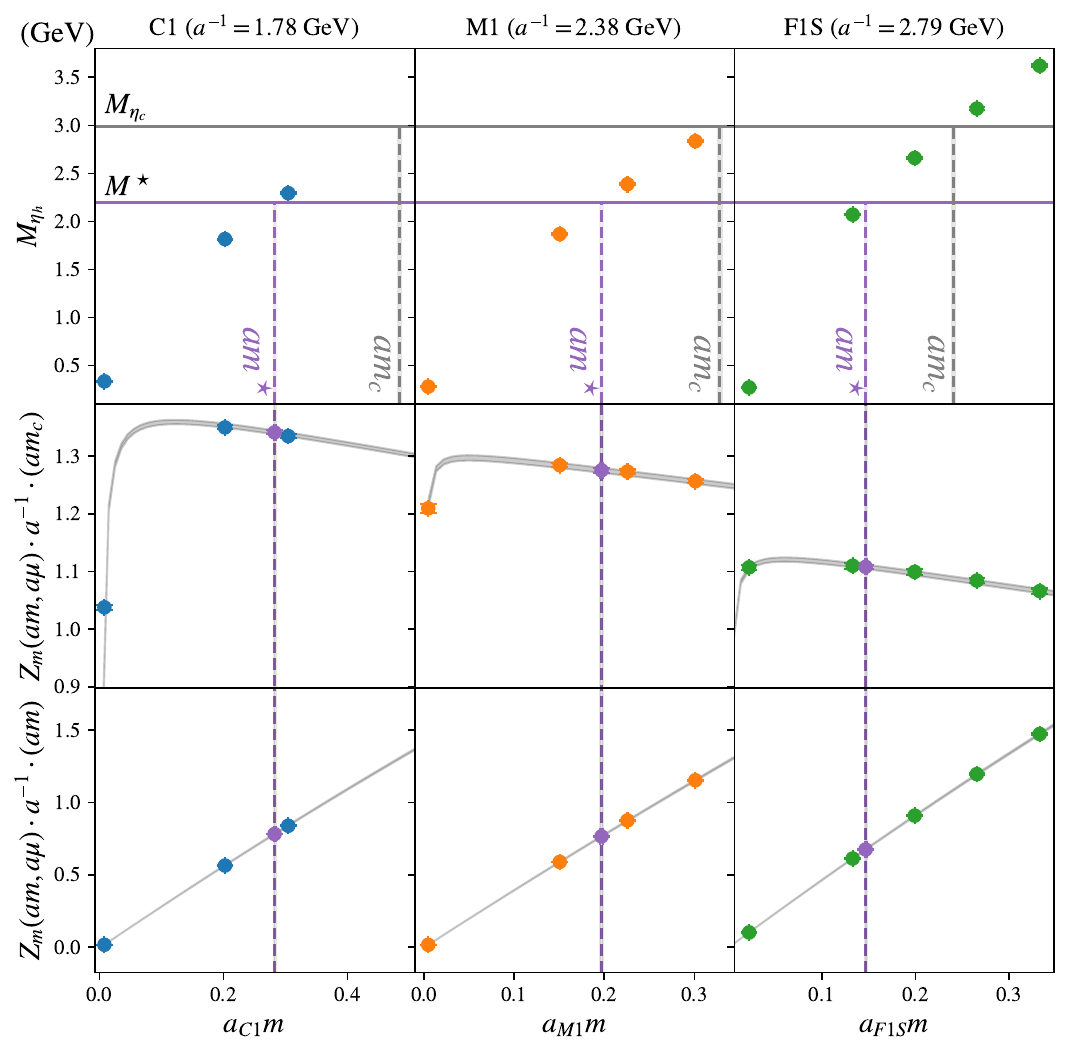}
    \captionsetup{belowskip=0pt}
    \vspace{-0.75\baselineskip}
    \caption{($\mu = 2$ GeV) Top row: $\eta_h$ meson mass $M_{\eta_h}$ vs bare mass $am$. $M_{\eta_c}^{\text{PDG}}$ corresponds to $am_c$ and choice of physical scale $M^\star$ corresponds to $am^\star$ on each lattice. Middle row: renormalised charm mass $Z_mm_c$ interpolated to $am^\star$. Bottom row: renormalised mass $Z_mm$ interpolated to $am^\star$.}
    \label{fig:interp}
\end{figure}
\begin{figure}
    \centering
    \begin{subfigure}{0.48\textwidth}
        \centering
        \includegraphics[height=2.2in]{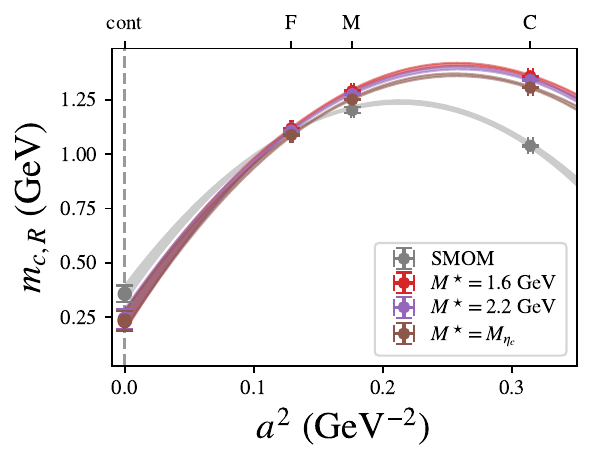}
        \vspace{-0.75\baselineskip}
        \caption{Continuum extrapolation of the renormalised charm mass in the massless scheme and in the massive scheme for various choices of $M^\star$.}
         \label{fig:m_c_ren}
    \end{subfigure}%
    \hfill
        \begin{subfigure}{0.48\textwidth}
        \centering
        \includegraphics[height=2.2in]{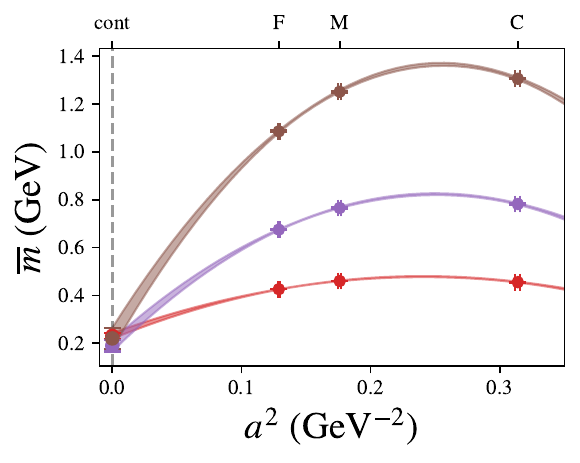}
        \vspace{-0.75\baselineskip}
        \caption{Continuum extrapolation of the renormalised mass for various choices of $M^\star$, corresponding to the scheme-defining renormalised mass $\overline{m}$.}
        \label{fig:mbar}
    \end{subfigure}
    \caption{Continuum extrapolations of $Z_mm_c$ and $Z_mm^\star$ using a quadratic ansatz in $a^2$. Renormalised charm mass from different schemes and choices of $M^\star$ should converge after all quantities are matched to a common scheme. Errorbars include statistical uncertainties and the error in lattice spacing only.}
    \label{fig:extrap}
    \vspace{-2mm}
\end{figure}
The methodology described above is implemented using three ensembles, and the preliminary results are shown in figures \ref{fig:interp} and \ref{fig:extrap}. Figure \ref{fig:interp} shows the procedure of referencing the $\eta_h$ meson mass measurements to calculate the bare charm mass on each lattice, and implementing the NPR procedure at a chosen physical scale $M^\star$. The top panel shows the interpolation performed for computing the bare charm mass in lattice units $am_c$ using the physical value of $M_{\eta_c}$. Additionally, the scale $M^\star$ is introduced, and again an interpolation is performed to find the corresponding bare mass in lattice units $am^\star$. Having applied these common scales across all lattice spacings, we can compute the renormalised charm mass using $Z_m(a\mu, am^\star)$, as shown in the middle panel. The same renormalisation constant is also used for renormalising the bare mass in the bottom panel. 

Having computed the renormalised charm and bare masses on all lattices for some choice of $M^\star$, we proceed to perform continuum extrapolations, as prescribed in equations (\ref{eq:m_c_ren}) and (\ref{eq:mbar}), to obtain the continuum quantities $m_{c,R}(\mu, M^\star)$ and $\overline{m}(\mu, M^\star)$. This procedure is shown in figures \ref{fig:m_c_ren} and \ref{fig:mbar} respectively. This is the extrapolation where we expect absorption of higher order cutoff effects with the use of the massive scheme. 

In figure \ref{fig:m_c_ren}, we see that higher order cutoff effects are evidently present in the use of both the massive and massless schemes. We observe an alleviation of $\mathcal{O}(a^4)$ effects with the use of the massive scheme, however this is not yet quantified using a variation in fit ansatze, or additional lattice spacings. The cutoff effects in these continuum extrapolations are further exacerbated by the small errors on the datapoints, which currently only include statistical uncertainties and the error in lattice spacing, and we plan to improve upon these with a careful study of systematic effects that arise in the several successive steps of interpolations in the procedure. The inclusion of these systematic errors in the massive scheme may also allow renormalised quantities to be canditates for a linear fit in $a^2$ for the continuum extrapolation, for some choices of $M^\star$.

%% file: 6_conclusions.tex

\section{Conclusions and future directions}
We have performed a preliminary numerical implementation of a massive momentum-subtraction scheme and used the renormalised charm-quark mass as a test case to study its effect on higher-order lattice artefacts. Qualitatively, employing this massive scheme, RI/mSMOM, helps reduce the severity of $O(a^4)$ cutoff effects in the continuum extrapolation, when compared to the same extrapolation using the massless scheme, RI/SMOM. In order to further investigate and quantify this claim, this study needs to be extended to include systematic uncertainties and fit variations. 

It is also of interest to do a systematic study of the dependence of this change in lattice artefacts with the choice of $\overline{m}$ (in practice, the choice of $M^\star$), as it could be expected that there would be an identifiable `window' of choices that are most effective in absorbing cutoff effects.

Note that the final step of matching to a common scheme still remains to be implemented, making use of one-loop perturbative calculations for $Z_m$ at finite quark mass (already computed in the original work \cite{Boyle:2016wis}). This step will also allow us to verify the convergence of all the schemes, and to produce a value of the renormalised charm-quark mass in a universal continuum scheme such as $\overline{\text{MS}}$ which can be directly compared to other computations.

If successful, this massive scheme could be used in the study of other bilinear operators, and the RI/mSMOM prescription could potentially be extended for four-quark operators as well.

%% file: 7_acknowledgements.tex

\section{Acknowledgements}

We thank Christopher Sachrajda, Andreas J\"{u}ttner, Peter Boyle, and Matteo Di Carlo for several useful discussions and inputs to this project. This work used the DiRAC Extreme Scaling service at the University of Edinburgh, operated by the Edinburgh Parallel Computing Centre on behalf of the STFC DiRAC HPC Facility (www.dirac.ac.uk) under project codes dp008 and dp207 using Grid and Hadrons. This equipment was funded by BEIS capital funding via STFC capital grant ST/R00238X/1 and STFC DiRAC Operations grant ST/R001006/1. DiRAC is part of the National e-Infrastructure. R.M. is supported by the Presidential Scholarship from the University of Southampton. F.E. has received funding from the European Union's Horizon Europe research and innovation programme under the Marie Sk\l{}odowska-Curie grant agreement No 101106913.